\documentclass{mem}
\usepackage{natbib}\usepackage{txfonts}\usepackage{balance}
\usepackage{graphicx}
\usepackage[a4paper]{hyperref}
\usepackage{epstopdf}
\begin{document}

\newcommand{\1}{\ensuremath{^{-1}}}
\newcommand{\2}{\ensuremath{^{-2}}}
\newcommand{\3}{\ensuremath{^{-3}}}
\newcommand{\4}{\ensuremath{^{-4}}}
\newcommand{\5}{\ensuremath{^{-5}}}
\newcommand{\6}{\ensuremath{^{-6}}}

%
%
\newcommand{\HZ}{\ensuremath{\,\mathrm{Hz}}}
\newcommand{\SEC}{\ensuremath{\,\mathrm{s}}}
\newcommand{\MIN}{\ensuremath{\,\mathrm{min}}}
\newcommand{\YR}{\ensuremath{\,\mathrm{yr}}}
\newcommand{\W}{\ensuremath{\,\mathrm{W}}}
\newcommand{\ERG}{\ensuremath{\,\mathrm{erg}}}
\newcommand{\EV}{\ensuremath{\,\mathrm{eV}}}
\newcommand{\KEV}{\ensuremath{\,\mathrm{keV}}}
\newcommand{\MEV}{\ensuremath{\,\mathrm{MeV}}}
\newcommand{\MIC}{\ensuremath{\,\mu\mathrm{m}}}
\newcommand{\CM}{\ensuremath{\,\mathrm{cm}}}
\newcommand{\KM}{\ensuremath{\,\mathrm{km}}}
\newcommand{\KMS}{\KM\SEC\1}
\newcommand{\PC}{\ensuremath{\,\mathrm{pc}}}
\newcommand{\KPC}{\ensuremath{\,\mathrm{kpc}}}
\newcommand{\MPC}{\ensuremath{\,\mathrm{Mpc}}}
\newcommand{\ARCSEC}{\ensuremath{\,\mathrm{arcsec}}}
\newcommand{\MAG}{\ensuremath{\,\mathrm{mag}}}
\newcommand{\K}{\ensuremath{\,\mathrm{K}}}
\newcommand{\Lo}{\ensuremath{\,\mathrm{L}_\odot}}
\newcommand{\Lsun}{\Lo}
\newcommand{\LsunH}{\ensuremath{\,\mathrm{L}_{\odot,H}}}
\newcommand{\Mo}{\ensuremath{\,\mathrm{M}_\odot}}
\newcommand{\Msun}{\Mo}

\newcommand{\chisq}{\ensuremath{\chi^2}}
\newcommand{\CHISQ}{\chisq}

%
%
\newcommand{\ML}{\ensuremath{\Upsilon}}
\newcommand{\MBH}{\ensuremath{M_\mathrm{BH}}}
\newcommand{\MBHO}{\ensuremath{M_\mathrm{BH,0}}}
\newcommand{\Lsph}{\ensuremath{L_\mathrm{sph}}}
\newcommand{\LEdd}{\ensuremath{L_\mathrm{Edd}}}
\newcommand{\wlLwl}{\ensuremath{\lambda L_\lambda}}
\newcommand{\LEddO}{\ensuremath{L_\mathrm{Edd,0}}}
\newcommand{\Msph}{\ensuremath{M_\mathrm{sph}}}
\newcommand{\sig}{\ensuremath{\sigma}}
\newcommand{\sige}{\ensuremath{\sigma_\mathrm{e}}}
\newcommand{\re}{\ensuremath{r_\mathrm{e}}}
\newcommand{\RBLR}{\ensuremath{R_\mathrm{BLR}}}

\newcommand{\sigmaT}{\ensuremath{\sigma_\mathrm{T}}}

\newcommand{\QH}{\ensuremath{Q(\mathrm{H})}}
\newcommand{\Q}[1]{\ensuremath{Q(\mathrm{#1})}}
\newcommand{\Te}{\ensuremath{T_\mathrm{e}}}
\newcommand{\X}[1]{\ensuremath{X(\mathrm{#1})}}
\newcommand{\REC}[2]{\ensuremath{\alpha(#1,#2)}}
\newcommand{\Lnu}{\ensuremath{L_\nu}}
\newcommand{\CS}[1]{\ensuremath{a_\nu(\mathrm{#1})}}
\newcommand{\TAUnu}{\ensuremath{\tau_\nu}}

\newcommand{\WL}{\ensuremath{\lambda}}
\newcommand{\forb}[2]{\mbox{[#1\,\textsc{\lowercase{#2}}]}} 
\newcommand{\perm}[2]{\mbox{#1\,\textsc{\lowercase{#2}}}} 
\newcommand{\WAVEA}[1]{\ensuremath{\WL\,#1\,\mbox{\rm\AA}}}
\newcommand{\WAVEM}[1]{\ensuremath{\WL\,#1\,\mathrm{\MIC}}}
\newcommand{\Ne}{\ensuremath{N_\mathrm{e}}}
\newcommand{\NH}{\ensuremath{N_\mathrm{H}}}
\newcommand{\N}[1]{\ensuremath{N(\mathrm{#1})}}
\newcommand{\AV}{\ensuremath{A_\mathrm{V}}}
\newcommand{\AWL}{\ensuremath{A_\WL}}
\newcommand{\EW}{\ensuremath{W_\WL}}
\newcommand{\HNU}{\ensuremath{h\nu}}
\newcommand{\Mdot}{\ensuremath{\dot{M}}}
\newcommand{\Z}{{\it z}}
\newcommand{\IONXI}{\ion{X}{+i}}
\newcommand{\STROM}{Str\"omgren}
%

%
%
\newcommand{\ten}[1]{\ensuremath{10^{#1}}}
\newcommand{\xten}[1]{\ensuremath{\times 10^{#1}}}
\newcommand{\parent}[1]{\ensuremath{\left(#1\right)}}
\newcommand{\parentB}[1]{\ensuremath{\left[#1\right]}}
\newcommand{\parfrac}[2]{\ensuremath{\left(\frac{#1}{#2}\right)}}
\newcommand{\parfracB}[2]{\ensuremath{\left[\frac{#1}{#2}\right]}}

\newcommand{\HB}{\ensuremath{\mathrm{H}\beta}}
\newcommand{\CIV}{\perm{C}{IV}}
\newcommand{\OIII}{\forb{O}{III}}
\newcommand{\wlLwlV}{\ensuremath{L_{5100}}}
\newcommand{\wlLwlUV}{\ensuremath{L_{1350}}}
\newcommand{\wlLwlMg}{\ensuremath{L_{3000}}}
\newcommand{\bolion}{\ensuremath{a}}
\newcommand{\bolV}{\ensuremath{b}}
\newcommand{\FWHB}{\ensuremath{V_{H\beta}}}
\newcommand{\FWCIV}{\ensuremath{V_\mathrm{CIV}}}
\newcommand{\FWMgII}{\ensuremath{V_\mathrm{MgII}}}
\newcommand{\sigCIV}{\ensuremath{\sigma_\mathrm{CIV}}}
\newcommand{\LHB}{\ensuremath{L_{H\beta}}}
\newcommand{\fvest}{\ensuremath{\tilde{f}}}
\newcommand{\gvest}{\ensuremath{\tilde{g}}}
\newcommand{\MgII}{\perm{Mg}{II}}
\newcommand{\lvirrad}{\mathcal{ML}}

\title{
Weighing black holes from zero to high redshift}

   \subtitle{}

\author{
A.~Marconi\inst{1} 
\and D.~Axon\inst{2}
\and R.~Maiolino\inst{3}
\and T.~Nagao\inst{4}
\and\\ P.~Pietrini\inst{1}
\and A.~Robinson\inst{2}
\and G.~Torricelli\inst{5}
          }


\institute{
Dipartimento di Astronomia e Scienza dello Spazio, Universit\'a di Firenze, Italy
\and
Physics Department, Rochester Institute of Technology, USA
\and
INAF -- Osservatorio Astronomico di Roma, Italy
\and
National Astronomical Observatory of Japan, Tokyo, Japan
\and
INAF -- Osservatorio Astrofisico di Arcetri, Firenze, Italy\\
}

\authorrunning{Marconi et al.}

\titlerunning{Weighin black holes}

\abstract{The application of the virial theorem provides a tool to estimate supermassive black hole (BH) masses in large samples of active galactic nuclei (AGN) with broad emission lines at all redshifts and luminosities, if the broad line region (BLR) is gravitationally bound. In this paper we discuss the importance of radiation forces on BLR clouds arising from the deposition of momentum by ionizing photons. Such radiation forces counteract gravitational ones and, if not taken into account, BH masses can be severely underestimated. We provide virial relations corrected for the effect of radiation pressure and we discuss their physical meaning and application. If these corrections to virial masses, calibrated with low luminosity objects, are extrapolated to high luminosities then the BLRs of most quasars might be gravitationally unbound. The importance of radiation forces in high luminosity objects must be thoroughly investigated to assess the reliability of quasar BH masses.

\keywords{Galaxies: nuclei --- Quasars: emission lines --- Galaxies: Seyfert}
}
\maketitle{}

\section{The BH mass ladder}

The discoveries of the tight correlations between the mass of BHs and luminosity, mass, velocity dispersion and surface brightness profile of their host spheroids have produced an enormous impact. \MBH-host galaxy relations allow a "census" of BHs in the local universe showing that local BHs were grown during AGN activity. Most of all, \MBH-host galaxy relations indicate that BHs are an essential element in the evolution of galaxies and the link between BH and host galaxy is probably caused by AGN feedback.
It is fundamental to measure BH masses in large samples of objects from zero to high redshifts to fully understand the effects of BH growth on galaxy evolution. Direct BH mass estimates based on spatially resolved stellar and gas kinematics are possible only in the local universe but can be the starting point to calibrate less direct techniques following a "BH mass ladder" \citep{peterson:2004a}.
Rung one of the BH mass ladder is provided by gas and stellar kinematical measurements resulting in the \MBH-\sige\ and \MBH-\Lsph\ relations (e.g.~\citealt{ferrarese:2005}).
Rung two is provided by reverberation mapping (RM) measurements; BH masses are given by $\MBH = f\Delta V^2\RBLR/G$, where \RBLR\ is the BLR average distance from the BH estimated with reverberation mapping, $\Delta V$ is the width of the broad emission line and $f$ is a scaling factor which depends on (unknown) BLR properties. Although this technique is potentially plagued by many unknown systematic errors, BH masses from RM are in agreement with the \MBH-\sige\ relation of normal galaxies.
This second rung has been calibrated by \cite{onken:2004} who estimated $f$ assuming that the AGN with RM observations follow the \MBH-\sige\ relation.
In principle this technique is not limited by distance but, in practice, RM observations of high-$z$, high-$L$ AGNs are exceedingly demanding in term of observing time.
The $\RBLR-L$ relation \citep{kaspi:2000} shows that continuum luminosity acts as a proxy for \RBLR\  leading to the so-called single epoch (SE) virial mass estimates. Rung three of the BH mass ladder has been calibrated, among others, by \cite{vestergaard:2006} starting from RM measurements. Since a single measurement can be wrong by up to a factor of~$\sim10$, SE \MBH\ are considered accurate only in a statistical sense and are routinely used to estimate \MBH\ in large sample of AGNs from zero to high redshifts.

\section{The effect of radiation pressure}

One of the basic assumptions of reverberation mapping is that the BLR is photoionized \citep{blandford:1982} implying that BLR clouds are subject to radiation forces arising from ionizing photon momentum deposition. Therefore BLR clouds will effectively "see" a smaller BH mass.
An order of magnitude estimate of this radiation force can be obtained with a simple heuristic model in which BLR clouds are test particles, optically thick to ionizing photons. The corrected "virial relation" is then
\begin{equation}
f V^2 = \frac{G}{r}\left(\MBH-\frac{a}{4\pi\, G\, c\, m_\mathrm{p}\, \NH}\right)
\label{eq:virial}
\end{equation}
where $r$ is the cloud distance from the AGN/BH, $m_\mathrm{p}$ is the proton mass, $a=L_\mathrm{ion}/L$ is the ionizing to total luminosity ratio and $\NH$ is the total cloud column density in the direction of the AGN/BH (see \citealt{marconi:2008}, hereafter M08, for more details).
The corrected mass estimator is 
\begin{equation}
\MBH=f\frac{V^2\, r}{G}+\frac{a}{4\pi\, G\, c\, m_\mathrm{p}\, \NH}\,L
\label{eq:gfactor}
\end{equation}
the first term is the "canonical" virial relation while the second one is the radiation force correction.
which can easily become relevant. For example, in AGNs with moderate column densities ($\NH\simeq\ten{23}$) and luminosities ($L\simeq \ten{11}\Lsun$), the correction becomes of the order of $2.7\xten{7}\Msun$, a typical value of the BH mass in a Seyfert galaxy (we have assumed $a=0.6$).
The above correction to \MBH\ is model dependent and \NH\ is unknown but
we can consider a virial estimator of the form:
\begin{equation}
\MBH=f\frac{V^2\, r}{G}+g\left[\frac{\wlLwl(5100\AA)}{\ten{44}\ERG\SEC\1}\right]
\end{equation}
One can then repeat the calibrations for rungs 2 and 3 as in \cite{onken:2004} and \cite{vestergaard:2006}.
The whole processes is presented and discussed in M08 and here we only outline the main results.
Calibration of Rung 2 is performed by imposing that the galaxies with RM data fall on the \MBH-\sige\ relation and the virial estimator for reverberation mapping observations (RM) is:
\begin{eqnarray}\label{eq:newRM}
\MBH& =& (3.1\pm1.4)\frac{V^2 R}{G}+\left(\ten{7.6\pm0.3}\wlLwlV\right)\Msun\nonumber
\end{eqnarray}
where $V^2 R/G$ is the virial product (in units of \Msun; \citealt{peterson:2004}) and \wlLwlV\ is \wlLwl\ at 5100~\AA\ (in units of $\ten{44}\ERG\SEC$).
The value $f\!=\!3.1$ here is almost a factor 2 smaller than $f\!=\!5.5$ found by \cite{onken:2004} due to the radiation pressure correction.
The $\sim\! 0.5$~dex r.m.s.\ scatter of BH masses around \MBH-\sige\ provides an indication of systematic errors affecting \MBH(RM) estimates compared to gas or stellar kinematical measurements.
The introduction of the $g$ parameter is not made to improve the fit (the scatter is not significantly reduced)
but is required by the physics of BLR clouds.
Our ability to determine an accurate empirical value of $g$ is limited, as were previous efforts to determine $f$, by size, composition and accuracy of the existing RM database. In particular it contains few sources with high Eddington ratios, which provide the tightest constraints on $g$. With this caveat in mind, however, we can jump onto the next rung of the BH mass ladder.
Calibration of Rung 3 is performed by imposing that SE \MBH\ estimates in galaxies with RM data are in agreement with the corresponding \MBH(RM)
and the virial estimator from single epoch (SE) observations is:
\begin{eqnarray}\label{eq:newSE}
\MBH/\Msun = \ten{6.13^{+0.15}_{-0.30}}\FWHB^2\wlLwlV^{0.5}+\ten{7.72^{+0.06}_{-0.05}}\wlLwlV\nonumber
\end{eqnarray}
where \FWHB\ is the FWHM of broad \HB\ (in units of 1000\KM\SEC). 
Without the radiation pressure correction, \MBH(SE)/\MBH(RM), has a r.m.s.~scatter of $\sim 0.4$ dex. Significantly, the scatter drops to $\sim 0.2$ dex when radiation pressure is taken into account indicating that SE masses are more accurate \MBH\ estimators than previously thought (see Fig.~2 of M08).
This $0.2$ dex r.m.s.\ scatter indicates only systematic errors made when using SE instead of RM masses. 
The physical meaning of this empirical $g$ value ($g\!\simeq\!\ten{7.7}\Msun$) can be assessed with our heuristic model. From Eq.~\ref{eq:gfactor}, the average \NH\ needed to obtain the empirical $g$ value is $\NH\!\simeq\!1.1\xten{23}\CM\2$ and is remarkably similar with the results from photoionization modeling studies of the BLR. Our empirical determination has indeed a simple physical meaning consistent with our current knowledge of the BLR.
A  further indication of the validity of our corrected virial relation is provided by Narrow Line Seyfert 1 galaxies.
These galaxies are believed to have small BHs compared to other AGNs and what is expected from the \MBH-\sige\ relation \citep{grupe:2004,zhou:2006}. Their small BH masses and high $L/\LEdd$ ratios suggest that these galaxies are rapidly growing their BHs.
With the radiation pressure correction, NLS1s move on to the \MBH-\sige\ relation, indicating that their BH masses were likely underestimated (see M08 for more details). 
\begin{figure*}[t!]
\centerline{
\includegraphics[clip=true,width=0.33\linewidth]{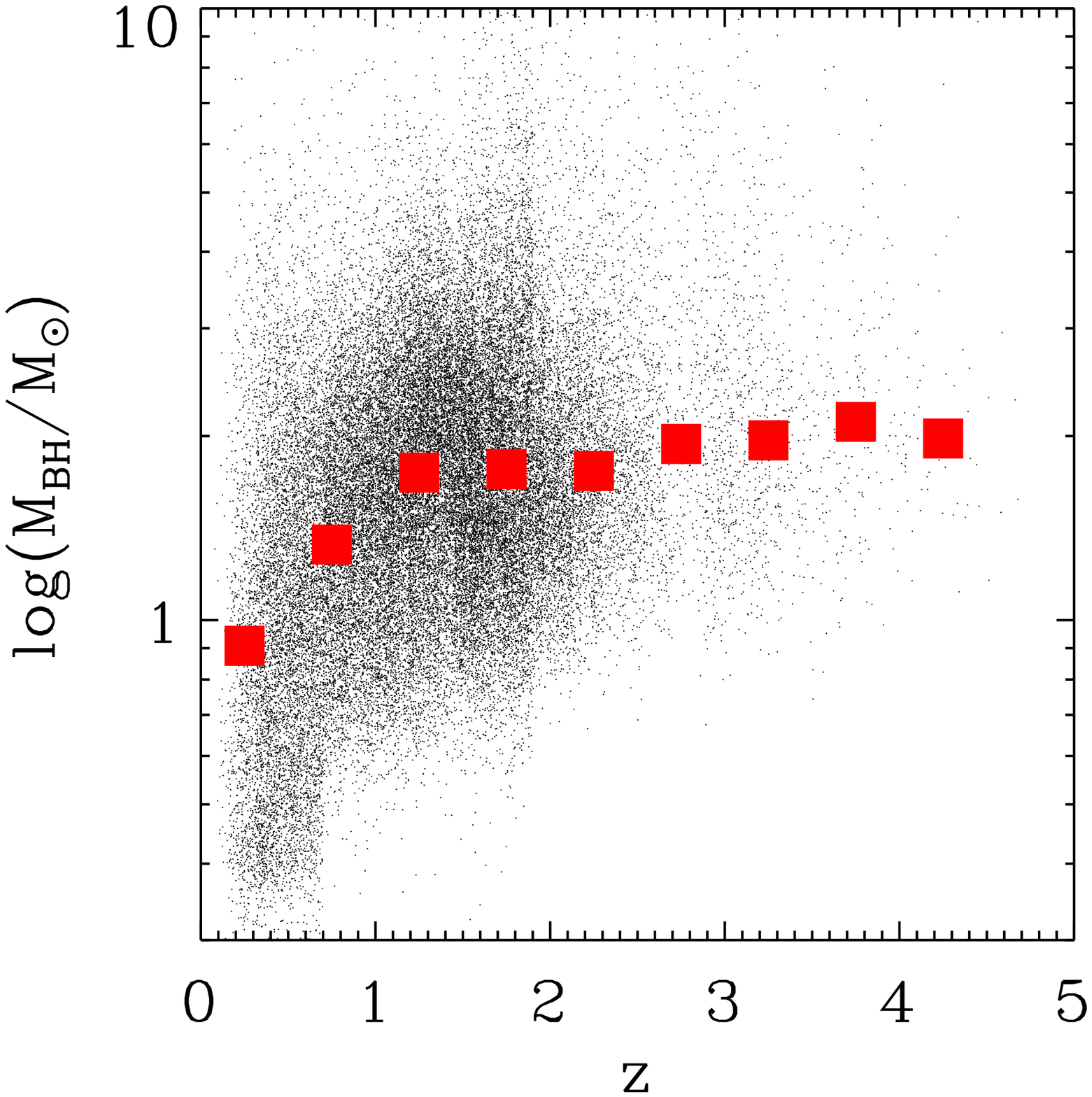} 
\includegraphics[clip=true,width=0.32\linewidth]{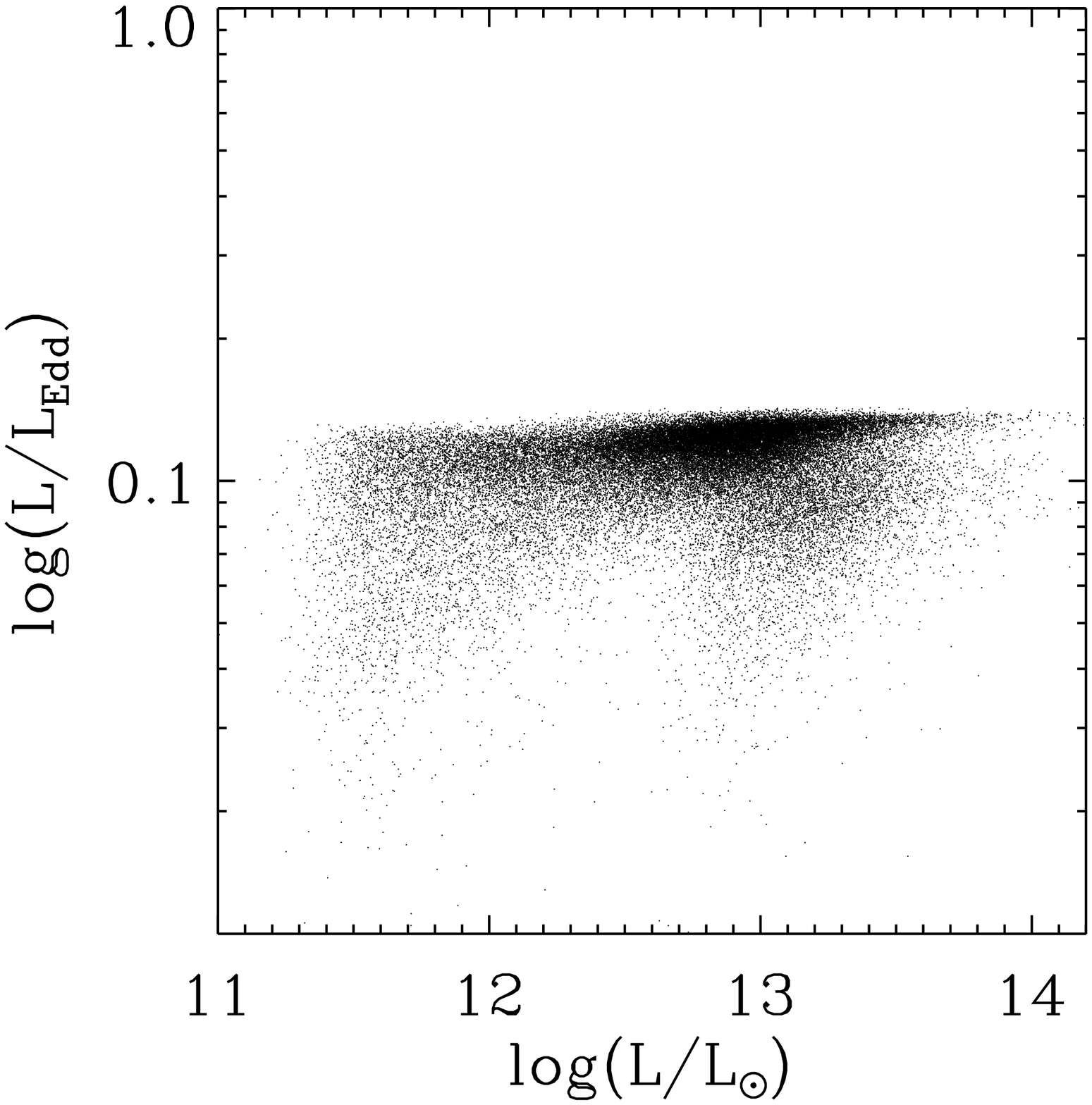}
\includegraphics[clip=true,width=0.33\linewidth]{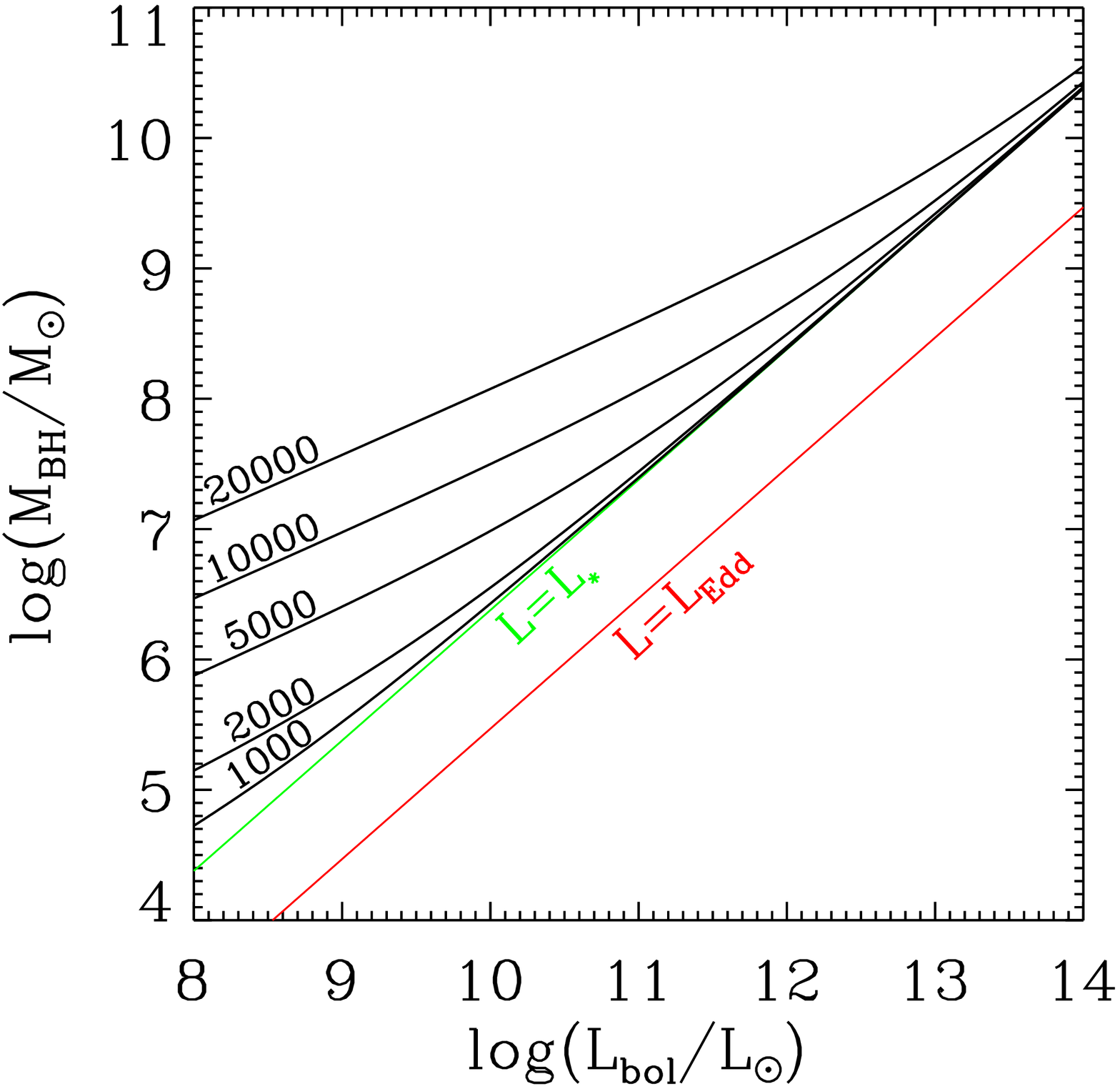}}
\caption{\footnotesize
Left: ratio between virial BH masses corrected (NEW) and non-corrected (OLD) for radiation pressure. Red filled squares denote average values in given redshift bins. Center: luminosity distribution of $L/\LEdd$ from NEW virial masses.
Right: Contours of constant line FWHM (solid black lines) in the $\MBH-L$  plane. Numbers denote FWHM values in \KM\SEC\1. The red and green lines denote the loci for $L=\LEdd$ and $L=L_\star$, where $L_\star$ is the critical luminosity at which the BLR is gravitationally unbound.
}
\label{fig:highL}\label{fig:scheme}
\end{figure*}

\section{High luminosity objects}

Virial relations involving UV lines are important to study sources at high redshift and it is possible to obtain virial relations for \CIV\ ($\lambda 1549\AA$) and \MgII\ ($\lambda 2798\AA$) which take into account radiation pressure like for \HB\ (Marconi et al.~2008, in preparation). Preliminary relations are:
\begin{eqnarray}\label{eq:newSE}
\MBH/\Msun &\simeq& \ten{6.5}\FWCIV^2\wlLwlUV^{0.5}+\ten{7.2}\wlLwlUV\nonumber\\
\MBH/\Msun &\simeq& \ten{6.1}\FWMgII^2\wlLwlMg^{0.5}+\ten{7.5}\wlLwlMg\nonumber
\end{eqnarray}
where $\FWCIV$, $\FWMgII$ are the FWHM of \CIV\ and \MgII\ broad lines respectively (in units of 1000\KM\SEC\1) and \wlLwlUV, \wlLwlMg\ are the continua at 1350 and 3000 \AA\ respectively (same notation as \wlLwlV).
The differences in $g$ factors w.r.t.~\HB\ depends on the different bolometric corrections.
To study the effect of radiation pressure corrected virial relations on high luminosity objects we consider the catalogue  by \cite{shen:2007} which contains line measurements for $\sim 60000$ SDSS quasars. In Fig.~\ref{fig:highL} we plot the ratio of new--to--old BH masses as a function with redshift (left) and the distribution of $L/\LEdd$ ratios as a function of luminosity (right). 
These two plots outline a few important points. BH masses corrected for radiation pressure are not necessarily larger than older ones; the average ratio ranges from $\sim 1$ to $\sim 2$ at high $z$ with a broad distribution of values. As shown above, the $f$ factor computed considering the radiation pressure correction is smaller by almost a factor 2; objects with low luminosities compared to their virial products will have a negligible radiation pressure correction and a BH mass which can be a factor $\sim 2$ smaller. At high $z$, sources in the catalogue have larger average luminosities while the distribution of line widths does not change appreciably with $L$ \citep{shen:2007}; hence the radiation pressure correction is more important and BH masses are, one average, larger.
The distribution of $L/\LEdd$ with luminosity is peculiar, reaching a well defined limiting value at about $L/\LEdd\simeq 0.15$.
Indeed when the luminosity is large the virial product becomes negligible with respect to the radiation correction. Hence $\MBH\propto L$ and the $L/\LEdd$ value saturates to a constant $L_\star/\LEdd$ value, where $L_\star$ is the critical luminosity which unbinds BLR clouds.
This behaviour is a clear consequence of the adopted correction to the virial relation (Eq.~\ref{eq:virial}). Moreover, such a relation is correct only if the right side is positive, i.e.~if the luminosity is smaller than $L_\star$ and the BLR is gravitationally bound, the fundamental underlying assumption which, of course, cannot be tested a-posteriori using the derived \MBH\ value.
Fig.~\ref{fig:highL}b clearly indicates that many high $L$ sources are close to the critical luminosity and this raises a potential problem.
If at quasar luminosities the BLR is experiencing only a small fraction, e.g.~$<10\%$, of the real gravitational potential then one should wonder whether the assumption of a gravitationally bound BLR is still acceptable.
In Fig.~\ref{fig:scheme}c we consider the $\MBH-L$ plane. For each point in this plane one can estimate the expected line width using
Eq.~\ref{eq:virial} and the $\RBLR-L$ relation. Black lines denote constant FWHM values while the red and green lines represent the $L=\LEdd$ (solid) and $L=L_\star$ loci.
The area between the red and green lines is populated by sources which are below Eddington but have their BLRs unbound.
Since most observed quasar spectra have $FWHM\!<\!10000\KM/\SEC\1$ \citep{shen:2007}, a quasar with $L\!>\!\ten{12}\Lsun$ will always have BLR clouds close to being unbound.
A similar situation characterizes NLS1 at lower luminosities ($1000\!<\!FWHM\!<\!2000\KM/\SEC$) which, in this picture, are low luminosity analogues of high luminosity quasars in terms of their BLR. The situation is different for Seyfert galaxies which populate a region where BLR clouds are well bound gravitationally.
This simple graph outlines a potential problem: the radiation force on BLR clouds in high $L$ quasars is extremely important compared to the gravitational attraction posing a potential problem on the validity of virial relations.
Of course, BLR clouds in high L quasars could be dominated by gravity if for instance the continuum is not isotropical and the BLR is illuminated by a much fainter continuum (e.g.~\citealt{proga:2008}), or BLR clouds in quasars might have much larger  column densities (e.g.~$\NH\!>\!\ten{25}\CM\2$) than in lower luminosity objects. Overall, it is clear that the radiation force is an important effect which should be taken into account in virial masses and that it is mandatory to assess whether the virial assumption of a gravitationally bound BLR is still valid in high luminosity quasars.


\end{document}